\documentclass[aps,prl,twocolumn]{revtex4}

\usepackage{amssymb}
\usepackage{graphicx}

\begin{document}

\title{Spurious trend switching phenomena in financial markets}
\author{V. Filimonov$^1$ and D. Sornette$^{1,2}$}
\affiliation{$^1$ ETH Zurich, Department of Management, Technology and Economics, Kreuzplatz 5, 8032 Zurich, Switzerland}
\affiliation{$^2$ Swiss Finance Institute, c/o University of Geneva, 40 blvd. Du Pont dÕArve
CH 1211 Geneva 4, Switzerland}
\date{\today}
\maketitle

{\bf Abstract}: The observation of power laws in the time to extrema of volatility, volume and intertrade times,
 from milliseconds to years, are shown to result
 straightforwardly from the selection of biased statistical subsets
of realizations in otherwise featureless processes such as random walks. The bias stems from the selection of
price peaks that imposes a condition on the statistics
of price change and of trade volumes that skew their distributions.
For the intertrade times, the extrema and power laws results from the format of transaction data.

\vskip 1cm
%=================================================================================

The random walk model of Bachelier (1900) \cite{Bachelier1900}, later extended into the geometrical
Brownian model (GBM) \cite{Cootner},  forms a reasonable first-order approximation of the dynamics of financial market prices.
The GBM constitutes the starting point for more refined modern models that take
into account the stylized facts documented in the last 50 years. While the GBM is based on the generally verified
absence of linear correlation of returns, many studies have show in addition the existence of 
long-memory in the volatility, volatility clustering and multifractality, fat tails in the distributions of returns, 
correlation between volatility and volume, time-reversal asymmetry, the leverage effect, gain-loss asymmetries and many others (see for instance,~\cite{cont,bouchaud}). 

Recently, Preis et al. \cite{preis1,preis2,preis3,preis4,preis5} have claimed the discovery of a new stylized fact
in the form of universal power laws associated with so-called switching points. They find that local maxima
of volatility and volume, and local minima of intertrade times, are reached and followed by power laws in the time to the extrema
that are reminiscent of critical points in Physics. The power laws are found to hold from time scales ranging from milliseconds to years.

Here, we show that these power laws are also found in the minimal random walk 
(they also hold as well for the GBM). They derive from the
statistical method used by Preis et al. to define the switching points. Using Occam razor, this suggests
that there is no need to invoke new properties for real financial prices, since essentially all characteristics
of switching points documented by Preis et al. \cite{preis1,preis2,preis3,preis4,preis5} are recovered in the GBM.
In other words, we show that the discovery of Preis et al. is likely an artifact of their statistical analysis, which
does not account for the impact of conditioning associated with the definition of switching points 
on the statistical properties of financial returns. Our finding applies directly to the volatility.  For volumes, 
one just needs to take into account the correlation between absolute price increments and volume, 
which is very strong at the daily time scale and weaker but yet pronounced at the tick-by-tick time scale. 
Concerning the dynamics of intertrade intervals, we show that it originates simply
from the format of the transaction price data.

\begin{figure}[h!]
    \includegraphics[width=\linewidth]{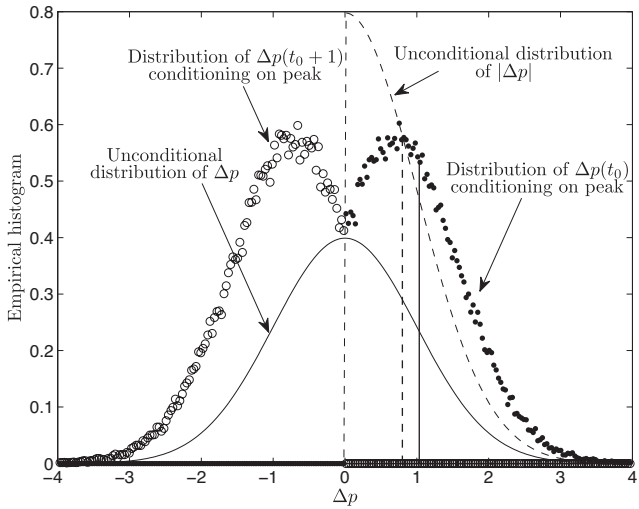}
   \caption{Conditional and unconditional distributions of price increments (see text). 
The vertical lines correspond to the unconditional expectation of $\Delta p(t)$ (dashed line) and to the expectation of 
$\Delta p(t_0)=p(t_0)-p(t_0-1)$ conditioned on the price peak occurring at time $t_0$ (continuous line) for a price
following the random walk model with iid Gaussian increments $N(0,1)$. The distribution of 
 $\Delta p(t_0)=p(t_0)-p(t_0-1)$ is strongly positively skewed, in particular with no negative realizations, 
 as results from the definition of $p(t_0)$. Similarly, the distribution of 
 $\Delta p(t_0+1)=p(t_0+1)-p(t_0)$ is strongly negatively skewed, in particular with no positive realizations, 
 as results from the definition of $p(t_0)$. } 
\label{fig_1}
\end{figure}

The local extrema considered by Preis et al.~\cite{preis1,preis2,preis3,preis4,preis5} are defined as follows.
The transaction price $p(t_0)$, where $t_0$ is a discrete time in the interval $[0,T]$ measured in transaction number or 
in calendar time, is defined to be a local maximum (resp. minimum) of order $\Delta t$ if there is no higher (resp. smaller) transaction price 
in the interval $t_0-\Delta t\leq t\leq t_0+\Delta t$. Independently of any assumption on the underlying 
generating process, this definition imposes stringent conditions on the price increments before and after 
the extrema.  For instance, by virtue of the definition of the existence of a local maximum, 
the conditional expectation of the last price increment $\Delta p(t_0)=p(t_0)-p(t_0-1)$ leading to the maximum
must be positively skewed  and  the conditional mean of $|\Delta p(t_0)|=|p(t_0)-p(t_0-1)|$
has to be larger than the unconditional mean of $|\Delta p(t)|$. This skewness also holds for earlier
increments from $t_0-2$ to $t_0-1$, from $t_0-3$ to $t_0-2$, and so on, with a decreasing amplitude
as one considers price increments further away from the peak. Similarly, the increment
$\Delta p(t_0+1)=p(t_0+1)-p(t_0)$ is negatively skewed, with the conditional mean being smaller than 
the unconditional mean. This negative skewness also holds for later increments with a progressively 
decreasing amplitude. While these effects can be studied in details analytically for the GBM, for pedagogy, 
we choose to show the result of numerical simulations for the random walk model
\begin{equation}\label{RW}
	 p(t)=\sum_{i=0}^{t-1}\xi(i)
\end{equation}
with Gaussian $N(0,1)$ iid increments $\xi(i)$ (fig.~\ref{fig_1}).
The same properties as shown in fig.~\ref{fig_1} hold for the volume and intertrade waiting times. 

Preis et al. \cite{preis1,preis2,preis3,preis4,preis5} have constructed the average trajectories
of volatility, volume and intertrade waiting times as a function of time to the extrema.
They have averaged twice, first, over different trend durations and, second, over different orders $\Delta t$.
A positive trend duration is simply defined as the time between the last
minimum to the next maximum.  A negative trend duration is defined as the time between the last maximum
to the next minimum. In order to be able to stack all positive and negative peaks respectively on each other,
Preis et al. have normalized each trend duration to $1$ by defining a dimensionless time $\varepsilon$
taking the value $0$ at the beginning of the trend and $1$ at the extrema.
They have found that volatility, volume and intertrade waiting times as a function of the 
 dimensionless reduced time $\varepsilon$ are power laws of the distance $|\varepsilon -1|$ to the extrema.

\begin{figure}[h!] 
\includegraphics[width=\linewidth]{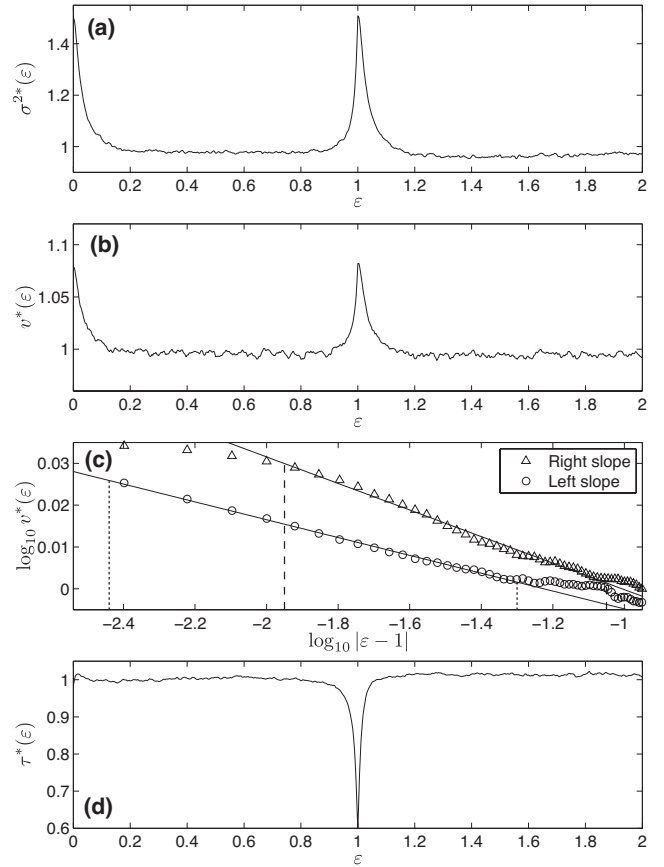}
\caption{(a) financial volatility $\sigma^{2*}(\varepsilon)$ obtained from  numerical simulations of the random walk model; 
(b) volume $v^*(\varepsilon)$ obtained from numerical simulations with the model of linear correlations between the volume and absolute price change (see text); (c) log-log plot of $v^*(\varepsilon)$ as a function of $|\varepsilon -1|$ on both side of the peak 
placed at $\varepsilon =1$. The two straight lines correspond to the power laws
$v^*(\varepsilon) \sim |\varepsilon -1|^{\beta_v}$, with $\beta_{v}^+=-0.16$ in
the range $10^{-1.95}<\varepsilon -1<10^{-1.05}$, and $\beta_{v}^-=-0.12$ in the range $10^{-2.45}<1-\varepsilon<10^{-1.3}$;
(d) intertrade time $\tau^*(\varepsilon)$ obtained from numerical simulations 
with the model with non-zero probability of ``walking the book'' (see text) with $p_0=0.5$.} \label{fig_2}
\end{figure}

We have performed exactly the same analysis as Preis et al. \cite{preis1,preis2,preis3,preis4,preis5} 
on the random walk model~(\ref{RW}) with Gaussian $N(0,1)$ iid increments $\xi(i)$
and find essentially the same power law dependences as found on empirical financial time series (see fig.~\ref{fig_2}).
Fig.~\ref{fig_2}a shows the average stacked peak of volatility as a function of $\varepsilon$ for the random walk model.
Note that the scale-free structure of the random walk ensures a weak dependence of this pattern on the order $\Delta t$,
as also found for the real financial data. The asymmetry of the peak, as
well as the two power-law decays around the peak, are found to hold also for the random walk model: 
a fit of the synthetic data generated with the random model with expression
\begin{equation}
	\sigma^{2*}(\varepsilon) \sim |\varepsilon -1|^{\beta_\sigma}
	\label{heynb}
\end{equation}
yields $\beta_{\sigma}^+=-0.16$ in
the range $10^{-1.95}<\varepsilon -1<10^{-1.05}$, and $\beta_{\sigma}^-=-0.12$ in the range $10^{-2.45}<1-\varepsilon<10^{-1.3}$.
Note that a better model for the singular behavior for $\varepsilon  \to 1$ amounts to replacing
(\ref{heynb}) by $\sigma^{2*}(\varepsilon) = a -b  |\varepsilon -1|^{\beta_\sigma}$ where $a$ is a constant,
so that the conditional volatility does not really diverge but exhibits a finite-time singularity characterized
by an infinite slope. A fit to the synthetic data usually gives $b>0$ and $\beta_\sigma \simeq 0.7$.
  
The asymmetry around the peak results from (i) the selection of realizations with peaks
that ensure that large positive (resp. negative) price changes are more probable before (resp. after) the peak and (ii) the asymmetry in dimensionless time $\varepsilon$, that is by definition fixed to the single trend for $0\leq\varepsilon\leq1$, but is free for $1\leq\varepsilon\leq2$ and may include in these values another trend, or even multiple trends, when $\Delta t$ is large. This effect becomes even stronger for tick-by-tick data, where the probability of having zero price change is high and the micro trend (of given order) is likely to be followed by a plateau of constant price.

These results are robust with respect to changes in the generating price process.
For instance, taking into account the existence of an atom (probability concentration) for zero price increments,
or of a negative auto-correlations of price increments at lag one due to bid-ask bounce and other effects
do not change qualitatively the results shown in Fig.~\ref{fig_2}. Only the values of the 
exponents $\beta_{\sigma}^+$ and $\beta_{\sigma}^-$ may be changed. 

Since the random walk model does not include any volume or intertrade intervals, we need to enrich it  slightly
to match the observations of Preis et al. \cite{preis1,preis2,preis3,preis4,preis5}. We start with the well-established empirical fact that volume is correlated with the absolute price change at many scales. Indeed, as Preis et al. have noted~\cite{preis1,preis2,preis3,preis4,preis5}, there are almost no correlation between signed price increments $\Delta p(t)$ and volume $v(t)$, due to the almost symmetric distribution of the $\Delta p(t)$. But, at the same time, it is well known that the correlation between absolute price increments $|\Delta p(t)|$ and volume is very strong
at the daily scale (see for instance the early review~\cite{Karpoff1987}) and weaker but clearly pronounced at the transactions scale. 
Here, it should be noted that studies of transaction data are not common because the transaction price is subjected to bid-ask bounce and thus does not reflect well market price moves. 

For illustration, let us take the example of transaction data of December'2011 futures on the DAX index over the period 07/11/2011--06/12/2011 (1'892'243 transactions in total) obtained from Bloomberg Historical Intraday Tick database. For this dataset, we find that the Pearson correlation coefficient between signed price increments $\Delta p(t)$ and volume $v(t)$ is equal to $1.02\%$ with the 95\% confidence interval 
$[0.88\%,1.16\%]$ clearly excluding the null hypothesis of a zero value. The correlation coefficient between absolute price increments $|\Delta p(t)|$ and volume $v(t)$ is more than 10 times stronger and equal to $15.7\%$ with its 95\% confidence interval equal to $[15.6\%,15.9\%]$.

Due to the correlation between volume and absolute price change,
the structure of the volume dynamics reproduces qualitatively that of the absolute price increments.
Assuming again that the price follows a random walk~(\ref{RW}), we account for the correlation between volume and absolute price change by specifying the following process for the volume: 
\begin{equation}
	v(t)=\Big| |\Delta p(t)|+\sigma_\mu\mu(t)\Big|,
	\label{hyjuym4u}
\end{equation}
 where $\mu(t)$ is an iid Gaussian noise $N(0,1)$ independent of $p(t)$ 
and $\sigma_\mu$ is a coefficient that controls the amplitude of the correlation 
between $|\Delta p(t)|$ and $v(t)$. Fig.~\ref{fig_2}b shows the 
average stacked volume dynamics $v^*(\varepsilon)$, conditional on 
price peak, as a function of $\varepsilon$ for the random walk model with (\ref{hyjuym4u})
and correlation coefficient of $0.2$. One can clearly observe
the same asymmetric peak as for the volatility shown in fig.~\ref{fig_2}a. 
Around the peak of volume, which coincides with that of price, fig.~\ref{fig_2}c
shows a log-log plot of the volume $v^*(\varepsilon)$ as a function of $|\varepsilon -1|$ on both side of the peak 
placed at $\varepsilon =1$. This log-log plot demonstrates the existence of two power laws, according to
$v^*(\varepsilon) \sim |\varepsilon -1|^{\beta_v}$, with $\beta_{v}^+=-0.16$ in
the range $10^{-1.95}<\varepsilon -1<10^{-1.05}$, and $\beta_{v}^-=-0.12$ in the range $10^{-2.45}<1-\varepsilon<10^{-1.3}$.
As for the conditional volatility around the peak, we note that a better model is 
$v^*(\varepsilon) = a' -b'  |\varepsilon -1|^{\beta_v}$ with $b'>0$ and $ \beta_v <1$.
 
As shown in fig.~\ref{fig_2}d, we are also able to reproduce the negative peak 
of the average intertrade intervals $\tau$ as a function of $\varepsilon$, conditioned on the existence
of a price peak, and its approximate power law dependence
in terms of $|\varepsilon -1|$ on both side of the peak. For this, 
we consider the same random walk model~(\ref{RW}) with discrete price increments $\xi(i)$ and we model intertrade 
intervals $\tau(t)$ as a mixture of (i) iid exponentially distributed random variables reflecting a Poisson process
for the order flow and (ii) an atom at $\tau=0$ with probability mass $p_0$, which accounts for those times $t_c$'s such that 
the price increment $\Delta p(t_c)$ has the same sign as the previous nonzero increment $\Delta p(t_c')$ 
(this means that the price moved in the same direction as before).
This is a simple toy model of the way that the transaction data is organized. In the order-driven exchanges such as 
the Chicago Mercantile Exchange and the Frankfurt Stock Exchange, which were analyzed by 
Preis et al.~\cite{preis1,preis2,preis3,preis4,preis5}, the matching of orders is performed according to a
real-time ``walk'' within the order book. When one submits a market order to buy or sell,
it is immediately executed at the best possible price. But when the submitted order is so large that
it could not be executed at one price (due to insufficient supply), a part of it is executed at the best next price 
and the rest -- at the second best next price, then at the third next best price -- and so on until the whole order is executed. 
This effect is known as ``walking the book''. These executions correspond to different transactions, since they 
are performed at different prices but, since they are triggered by the 
same order, all of them have identical time stamp
in  the transaction log-file. At the same time, such sequences of transactions are moving
price significantly, especially if more than two levels in the order book are involved. Therefore, the logfile of 
transaction records a finite fraction of trades with non-zero price changes and zero intertrade time interval.
As demonstrated by fig.~\ref{fig_2}d,
this effect is sufficient to explain the negative peak in the time series of the intertrade trades, reported
initially by Preis et al.~\cite{preis1,preis2,preis3,preis4,preis5}.

\begin{figure}[h!] 
\includegraphics[width=\linewidth]{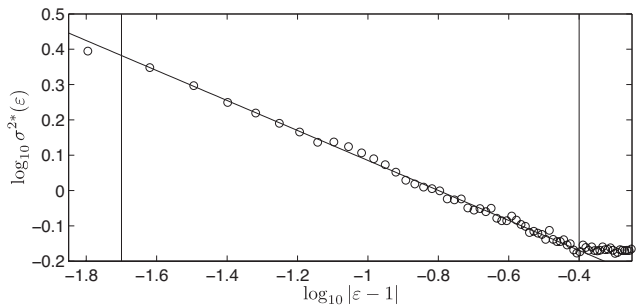}
\caption{Log-log plot of the volatility $\sigma^{2*}(\varepsilon)$ obtained from numerical simulations of the quasi-multifractal model~(\ref{qmf}-\ref{qmf_B}) introduced in Ref.~\cite{Saisor,Saifili}  with parameters $\varphi=0.1$ and $\sigma^2=5$. The straight line corresponds to the power law with the exponent $\beta_{\sigma}=-0.45$.} \label{fig_3}
\end{figure}

Finally we would like to notice that the range of the apparent power laws and their exponents obtained in the random walk model~(\ref{RW}) do not match quantitatively the results reported by Preis et al. \cite{preis1,preis2,preis3,preis4,preis5}. However, supplementing the random walk model with another stylized fact of real financial time series, it is possible to reproduce precisely both the quantitative values of the exponents and the range of scales over which the power laws hold. To illustrate this result, we will use the quasi-multifractal process \cite{Saisor,Saifili} that accounts for the heavy tails of price increments and  the long memory of absolute returns. This model represents price $p(t)$ as
\begin{equation}\label{qmf}
	p(t)=\sum_{i=0}^{t-1}\xi(i)e^{\omega(i)},
\end{equation}
where $\xi(i)$ are iid Gaussian $N(0,1)$ variables and $\omega(i)$ are independent of $\xi(i)$ Gaussian random variables with zero mean and covariance matrix
\begin{equation}\label{qmf_B}
	\mathbb{B}_{i,j}=%\frac{\sigma^2}4C(|i-j|),~
	\frac{\sigma^2\varphi}2\int_0^\infty\frac{dx}{\big((1+x)(1+x+|i-j|)\big)^{\varphi+1/2}}~.
\end{equation}
Fig.~\ref{fig_3} presents an example of the simulations of the quasi-multifractal model~(\ref{qmf}-\ref{qmf_B}) for values of 
the parameters ($\varphi=0.1$ and $\sigma^2=5$) that replicate the power laws found in the volatility patterns of 
the S\&P 500 stocks documented in Ref.~\cite{preis1,preis3}, with exponents $\beta_{\sigma}=-0.46$ in the range $10^{-1.7}<\varepsilon -1<10^{-0.4}$.

In summary, we have shown that the definition of price peaks imposes a condition on the statistics
of price change and of trade volumes that skew their distributions sufficiently
to explain the occurrence of power laws in the time to these peaks, even in the simplest possible
model, the random walk. Though the minimal random walk model could not reproduce precisely the exact exponents and ranges of power laws reported by Preis et al. \cite{preis1,preis2,preis3,preis4,preis5}, more elaborated models allow to match these values quantitatively. This statement was illustrated with the quasi-multifractal model~\cite{Saisor,Saifili} that accounts for the long-term memory and heavy tailed statistics of real price returns. For the intertrade times, we have shown that the extrema and power laws results from the format of transaction data. 
We are thus led to conclude that there is no new ``switching'' phenomenon, 
as the power laws are straightforward consequences of the selection of biased statistical subsets
of realizations in otherwise featureless processes.  In the switching phenomena
reported by Preis et al. \cite{preis1,preis2,preis3,preis4,preis5}, 
there is no more than statistical conditioning and some correlations.

We are grateful to George Harras for useful discussions while preparing this manuscript.

%=================================================================================

\end{document}